\documentclass[pss]{wiley2sp}
\usepackage{amsmath}

\pdfoutput=1
\definecolor{dark-blue}{rgb}{.0,0.0,.75}
\definecolor{dark-green}{rgb}{.0,0.75,.0}
\usepackage{ifpdf}
\ifpdf
\usepackage[pdftex,hyperindex,
	hypertexnames=false,
	bookmarksnumbered=true,
	colorlinks=true,
	pdfstartview=FitV,
	linkcolor=blue,
	citecolor=dark-green,
	urlcolor=dark-blue,
            ]{hyperref}
\else
\fi
\usepackage{cite}
\newcommand{\ddt}{\frac{\partial}{\partial t}}
\newcommand{\ihddt}{i\hbar\ddt}

\newcommand{\qav}[1]{\left \langle #1 \right \rangle}             %% Quantum mechanical AVerage...
\newcommand{\bra}[1]{\left < #1 \right |}               %BRAket
\newcommand{\ket}[1]{\left | #1 \right >}               %braKET
\newcommand{\pa}[1]{\left ( #1 \right )}                %PArenthesis
\newcommand{\eqw}[2]{\begin{equation}\\ \label{#1}\\ #2\\ \end{equation}}       %EQuation With label
\newcommand{\eq}[1]{\begin{equation}\\#1\\ \end{equation}}                      %EQuation
\newcommand{\eqa}[1]{\begin{eqnarray}#1 \end{eqnarray}}                         %EQuation Array
\newcommand{\bi}{\begin{itemize}}
\newcommand{\ei}{\end{itemize}}
\newcommand{\be}{\begin{enumerate}}
\newcommand{\ee}{\end{enumerate}}
\newcommand{\reff}[1]{Fig.~\ref{#1}}
\newcommand{\refe}[1]{Eq.~(\ref{#1})}

\newcommand{\mrm}[1]{\mathrm{#1}}

\begin{document}

\title{Electron-electron Relaxation in Disordered Interacting Systems}

\titlerunning{Electron-electron Relaxation in Disordered Interacting Systems}

\author{
  Peter Bozsoki\textsuperscript{\textsf{\bfseries 1,\Ast}},
  Imre Varga\textsuperscript{\textsf{\bfseries 2}},
  Henning Schomerus\textsuperscript{\textsf{\bfseries 1}}}

\authorrunning{P.~Bozsoki et al.}

\mail{e-mail
  \textsf{p.bozsoki@lancaster.ac.uk}, Phone
  +44-1524-593291, Fax +44-1524-844037}

\institute{
  \textsuperscript{1}\,Department of Physics, Lancaster University, UK-LA1 4YB
  Lancaster, United Kingdom\\
  \textsuperscript{2}\,Elm\'eleti Fizika Tansz\'ek, Fizikai Int\'ezet, Budapesti
  M\H{u}szaki \'es Gazdas\'agtudom\'anyi Egyetem, Budafoki \'ut 8, H-1111
  Budapest, Hungary}

\received{XXXX, revised XXXX, accepted XXXX}
\published{XXXX}


\abstract{
\abstcol{
We study the relaxation of a non-equilibrium carrier distribution under the
influence of the electron-electron interaction in the presence of disorder.

Based on the Anderson model, our Hamiltonian is composed from a single
particle part including the disorder and a two-particle part accounting for
the Coulomb interaction. We apply the equation-of-motion approach for the
density matrix, which provides a fully microscopic description of the
relaxation.
}{
Our results show that the nonequlibrium distribution in this closed and
internally interacting system relaxes exponentially fast during the initial
dynamics. This fast relaxation can be described by a phenomenological
damping rate. The total single particle energy decreases in the
redistribution process, keeping the total energy of the system fixed. It
turns out that the relaxation rate decreases with increasing
disorder.
}}
\maketitle

\section{Introduction}

Understanding the interplay of disorder and interaction constitutes a
challenging problem in condensed matter physics \cite{reviews}. Even weak
interactions cause {a considerable} amount of complexity \cite{Basko}. One
of the most intriguing phenomena, the phase relaxation due to interaction,
has already been treated in the presence of disorder \cite{Gersh}. In many
cases, however, either of the two major ingredients, the disorder and the
inter-particle interaction are taken {into account} only perturbatively. In
numerical simulations this type of approximation is not necessary. However,
finite-size limitations do affect the results, even though further
controlled simplifications may nevertheless be necessary to impose.

In this paper we present a case study where we have investigated how the
interplay of disorder and long-range interaction {affects} the relaxation
of a non-equilibrium energy distribution of electrons in one
dimension.

 In this paper our aims are twofold. On the one hand, we want to determine
whether the relaxation within a closed interacting system can be described
by a phenomenological damping rate. On the other hand, we perform the first
steps in the direction of constructing a theory of relaxation treating both
disorder and interaction on an equal microscopic footing. Our approach is
based on the equation of motion technique applied on elements of the
density matrix. This technique has achieved a great success in numerous
applications in semiconductor optics for systems with strong many-body
correlations \cite{PQE, CSO}. Our experience shows that the interplay of
disorder and interaction does yield novel and interesting phenomena
\cite{Bozsoki:PRL06, Bozsoki:JLumi07}. Here we provide one more evidence
that it can also be used to tackle challenging problems in the field of
interacting disordered systems.

The phase relaxation due to inter-particle interactions in disordered
systems has already been investigated analytically, for a recent review,
see \cite{Gersh}. As a summary, the presence of weak disorder accelerates
the relaxation process, since the scattering probabilities increase due to
the lack of momentum selection rules an ordered system would impose. The
question of strong disorder, however, is not yet fully understood. A first
step using numerical simulations was performed in \cite{Varga03}, where a
Boltzmann-type equation was solved with an initial condition of a
non-equilibrium carrier distribution, {which relaxes} over the course of
time towards a Fermi-Dirac-type distribution. While the analytical
prediction for weak disorder was recovered in the case when short-ranged
interactions were allowed between the particles, for both long ranged and
short ranged interaction the strong disorder limit yielded a decrease in
the relaxation rate. The latter behavior was attributed to the small size
of the localization volume of the one particle-states, resulting
in smaller scattering probabilities.

Since the previous work of I.~Varga \textit{et~al.}~\cite{Varga03} suffered
from several simplifying assumptions, in the present work we perform
computations along similar ideas, but using a much better founded
formalism, namely the equation of motion of the density matrix
\cite{CSO,Bozsoki:PRL06,Bozsoki:JLumi07}.

\section{Model}

\subsection{The Hamiltonian}
Our model is based on a one dimensional tight--binding model with energetic
disorder, represented by the energies of the sites denoted by
$\varepsilon_j$ for the $j^\mrm{th}$ site \cite{CSO,Bozsoki:JLumi07}. The
model also includes nearest neighbor hopping {matrix elements} $J_{ij} = J$
between sites $i$ and $j$. Using a site basis where $\ket{i}$ denotes the
basis function at site $i$ it is straightforward to formulate the
non-interacting part of the Hamiltonian:
\begin{equation}
\label{eq:model:Hamiltonian-eh}
H_0 = \sum_i \varepsilon_i \ket{i} \bra{i}
                + \sum_{<ij>} J_{ij} \ket{i} \bra{j} \, ,
\end{equation}
where $<ij>$ corresponds to a sum limited to the nearest neighbors.

The term responsible for the two-body interaction reads \cite{CSO}
\eqw{eq:model:ham-Coul}{
  H_{\mrm{C}} =  \frac12 \sum_{i,j} V_{ij} \,
                         \ket{i} \ket{j} \bra{j} \bra{i} \, ,
}
where only monopole--monopole terms have been included. In
\refe{eq:model:ham-Coul}, the sum represents the repulsive Coulomb
interaction among charge carriers within a single band. The regularized
Coulomb matrix element considered here is given by
\begin{equation}
  V_{ij} = \frac{U_0}{|i-j|a + a_0} ,
\label{eq:model:Coulomb}
\end{equation}
with a positive constant $U_0$. Here $a = 5$~nm is the site separation
\cite{CSO} and the term $a_0=0.5\,a$ removes
the unphysical singularity of the lowest excitonic bound state arising from
the restriction to monopole--monopole terms in one dimension~\cite{CSO,
Bozsoki:JLumi07}.

\subsection{Single Particle Basis}
In order to derive the equations of motion it is useful to write the
Hamiltonian in second-quantized formalism. As we are interested in the
dynamics of single particles, we transform the Hamiltonian into the
eigenbasis of single particles via diagonalizing $H_0$ in
\refe{eq:model:Hamiltonian-eh}
\eq{
H_0^\mrm{SP} = \sum_{\alpha} \epsilon_{\alpha}c^{\dag}_{\alpha} c_{\alpha} \,.
}
After the transformations the Hamiltonian containing the Coulomb interaction
reads as
\eq{
H_C^\mrm{SP} = \frac12 \sum_{\alpha \beta \atop \alpha' \beta'}
V_{\alpha\beta}^{\alpha'\beta'} c_{\alpha'}^\dag c_{\beta'}^\dag c_{\beta} c_{\alpha}
}
with the transformed matrix elements
\eq{
V_{\alpha\beta}^{\alpha'\beta'} =
      \sum_{ij} \Phi^*_{\alpha'}(i)\Phi^*_{\beta'}(j)
          V_{ij}\Phi_{\beta}(j)\Phi_{\alpha}(i) \,.
}
{Here} $\Phi_{\alpha}(i)$ is the projection of the electronic eigenstate with
quantum number $\alpha$ onto site $j$.

\section{The equation of motion}

The electron population in an eigenstate $\alpha$ is given by the
expectation value of the number operator:
\begin{equation}
n_\alpha=\qav{c_\alpha^\dag c_\alpha} \,.
\end{equation}
The off--diagonal extensions
\eq{
n_{\alpha\beta}=\qav{c_\alpha^\dag c_\beta}
}
are called coherences \cite{CSO} and play an important role in the present
calculation. The dynamics of all these matrix elements of the density matrix
are coupled via the Coulomb interaction. Thus one has to calculate the time
evolution of the full density matrix, which leads to the following equation
of motion:
\eqa{ \label{eq:EoM_n_full}
\ihddt n_{12} &=& \pa{\epsilon_2 - \epsilon_1 } n_{12} \nonumber \\
&& - \sum_{\alpha \beta \gamma}
	\pa{ U_{\alpha \beta}^{2 \gamma} C_{1 \gamma \beta \alpha}
			- U_{1 \beta}^{\alpha \gamma} C_{\alpha \gamma \beta 2} } \,,
}
with the correlated two-particle quantity being
\begin{equation}
C_{\alpha \beta \gamma \delta} = \qav{c_\alpha^\dag c_\beta^\dag c_\gamma c_\delta} \,,
\end{equation}
and the symmetrized Coulomb matrix element defined as
\begin{equation}
 U_{\alpha \beta}^{\alpha' \beta'} = \frac12
       \pa{ V_{\alpha \beta}^{\alpha' \beta'}
		 -  V_{\beta \alpha}^{\alpha' \beta'} } \,.
\end{equation}
Note that in \refe{eq:EoM_n_full} the indices $1,2$ denote general
single--particle states. They were introduced only to help the reader to
separate the summing and non-summing indices within the equation. Due to
the coupling of $n_{\alpha \beta}$ to $C_{\alpha \beta \gamma \delta}$ in
principle the equation of motion for the latter quantity should be derived
as well, and the two coupled equations should be solved together. It is,
however, possible to separate the contribution of uncorrelated particles
and pure two-particle quantum correlation with the help of {a} factorization
scheme, which yields \cite{PQE}
\begin{equation} \label{eq:dC_def}
C_{\alpha \beta \gamma \delta} =
  n_{\alpha \delta} n_{\beta\gamma} - n_{\alpha \gamma} n_{\beta \delta}
     - \Delta C_{\alpha \beta \gamma \delta} \,.
\end{equation}
Assuming the contribution of the pure two--particle part to be negligible in
comparison with the uncorrelated terms, i.e.,
\begin{equation}
\Delta C_{\alpha \beta\gamma \delta} \equiv 0 \, ,
\end{equation}
we end up with {the equation of motion}
\begin{eqnarray} \label{eq:EoM_n}
\ihddt n_{12} &=& \pa{\epsilon_2 - \epsilon_1 } n_{12} \nonumber \\
&& +	\sum_{\alpha \beta \gamma}
	U_{\alpha \beta}^{2 \gamma}
	\pa{n_{1 \alpha} n_{\gamma \beta} - n_{1 \beta} n_{\gamma \alpha}}
	\nonumber \\
&& - \sum_{\alpha \beta \gamma}
	U_{1 \beta}^{\alpha \gamma}
	\pa{n_{\alpha 2} n_{\gamma \beta} - n_{\alpha \beta} n_{\gamma 2}}
\end{eqnarray}
for the populations and the coupled off-diagonal elements.

\section{Results}

In order to solve \refe{eq:EoM_n} using standard numerical integration
techniques, we start with an initial population distribution
\begin{equation}
n_\alpha = \frac{1}{Z}e^{-(\epsilon_\alpha-E_c)^2/w^2} \ ,
\end{equation}
which is centered around the energy $E_c$ with a half width {at} full
maximum $w$. $Z$ is fixed by the total number of electrons $\sum_\alpha
n_\alpha=N/2$, where $N$ is the number of sites, chosen to be $N=20$. The
width of the distribution is chosen to be $w=J$ in order to avoid unphysical
cases of too large values of $n_\alpha$.

\begin{figure}
  \includegraphics*[width=.75\linewidth,height=\linewidth,angle=-90]{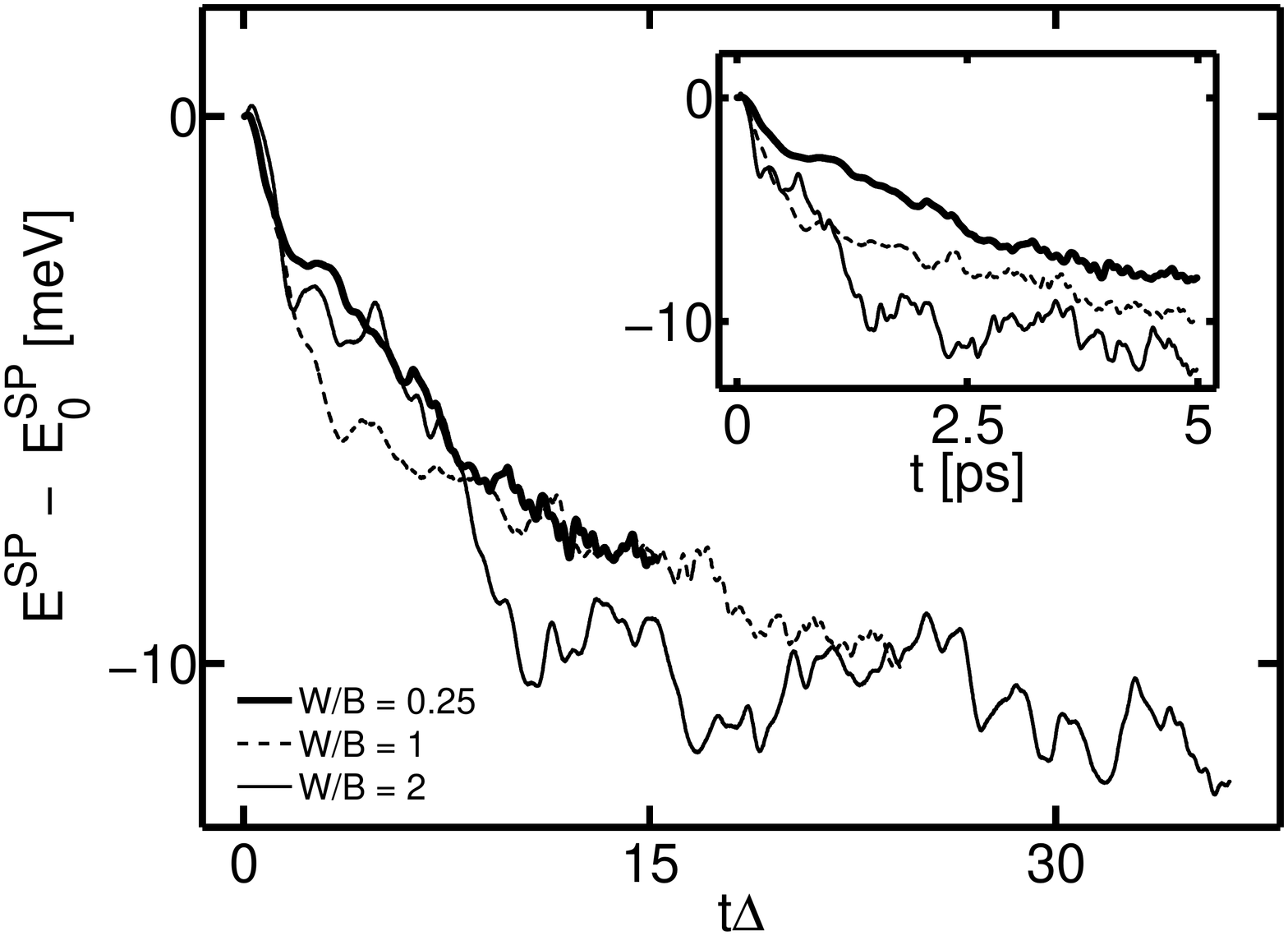}
  \caption{Total single particle energy as a function of time. The inset
					 shows the raw data, in the main panel time is rescaled with the
						mean level spacing.
 \label{fig:totalEnergy}}
\end{figure}

In order to quantify how much the system has relaxed we use two physical
quantities. One of them is the total single particle energy
\begin{equation}
\label{eq:Esp}
E^\mrm{SP} (t) = \sum_\alpha \epsilon_\alpha n_\alpha(t) \ ,
\end{equation}
which evolves in time as $n_\alpha$ changes. The other quantity is a
measure which tells us how far the system is from equilibrium. It is
defined by the root mean square of standard deviation of a
nonlinear fit to a Fermi-Dirac distribution:
\begin{equation}
\label{eq:sigma}
\sigma(t) = \left[ \sum_\alpha \pa{n_\alpha(t)
             - n_\mrm{FD}\pa{\epsilon_\alpha}}^2 \right] ^{1/2} \, .
\end{equation}
In general \refe{eq:EoM_n} can be solved only numerically for a particular
realization of the disorder. Thus an averaging over numerous realizations
has to be performed in order to
{eliminate the dependence on the specific} disordered potential
landscape. Applying this to the total energy yields the results shown in
\reff{fig:totalEnergy}.
The total single particle energy decreases from its initial value with a
rate depending on the mean level spacing of single-particle states $\Delta
\equiv (W + B)/N/\hbar$ ($\hbar = 0.658$~meVps) that results in a roughly
universal behavior when time is rescaled accordingly.

The configuration average is possible only for quantities like the single
particle energy, $E^\mrm{SP}$ or the {standard deviation}, $\sigma$.
{On the other
hand, $n_{\alpha \beta}$ corresponds to the eigenstates ${\alpha}$ and
${\beta}$ and the related eigenenergies, which are strongly dependent
on the particular disorder realization. I.e.,  even the same $\alpha$ and
$\beta$ refer to different eigenstates, which leads to
non-comparable $n_{\alpha\beta}$ for different realizations.}

Now let us turn towards the evaluation of the relaxation rate, $\Gamma$, which
is defined via the assumption that $\sigma(t)$ decreases exponentially in the
short-time {regime} of the dynamics, i.e.,
\begin{equation}
\label{eq:sigmat}
\sigma(t) = \sigma_0 e^{-\Gamma t} \, .
\end{equation}
\begin{figure}
  \includegraphics*[width=.75\linewidth,height=\linewidth,angle=-90]{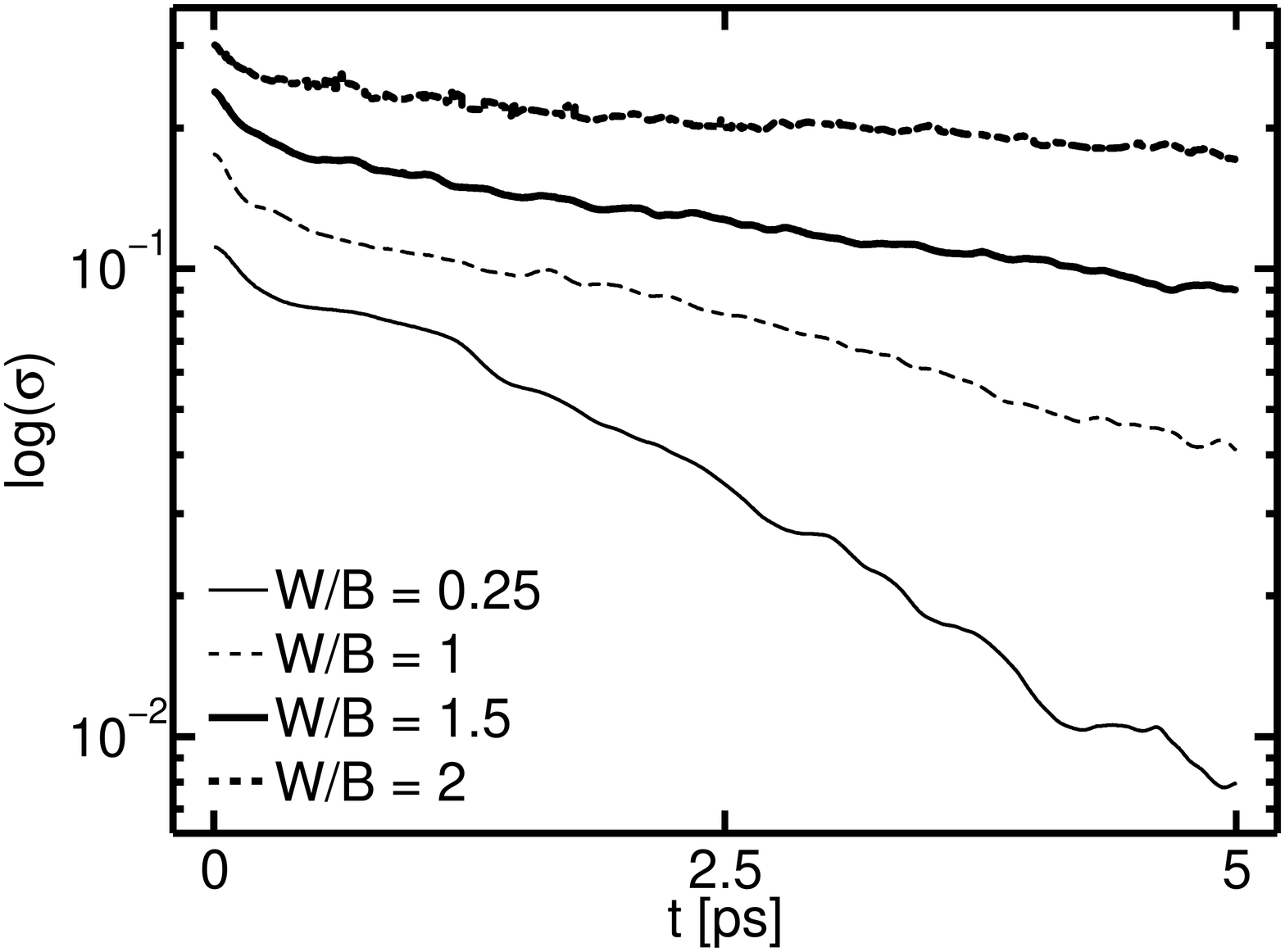}
  \caption{Error to a fit to a Fermi--Dirac distribution for a fixed interaction
           strength, $U_0=B=32$~meV for several values of disorder strength.
\label{fig:sigma}}
\end{figure}
Indeed, our assumption is corroborated by the numerical simulations,
shown in \reff{fig:sigma}. The figure shows the fall--off of the
{deviation} $\sigma$ as a function of time on a semi-log plot for
several values of disorder in units of the bare bandwidth, $B=4J$. [As we
have fixed the value of $J=8$~meV the bare bandwidth is $B=32$~meV. Hence
the timestep for the integration in \refe{eq:EoM_n} was chosen to be
$\delta t\approx 1$~fs.]

\begin{figure}
  \includegraphics*[width=.75\linewidth,height=\linewidth,angle=-90]{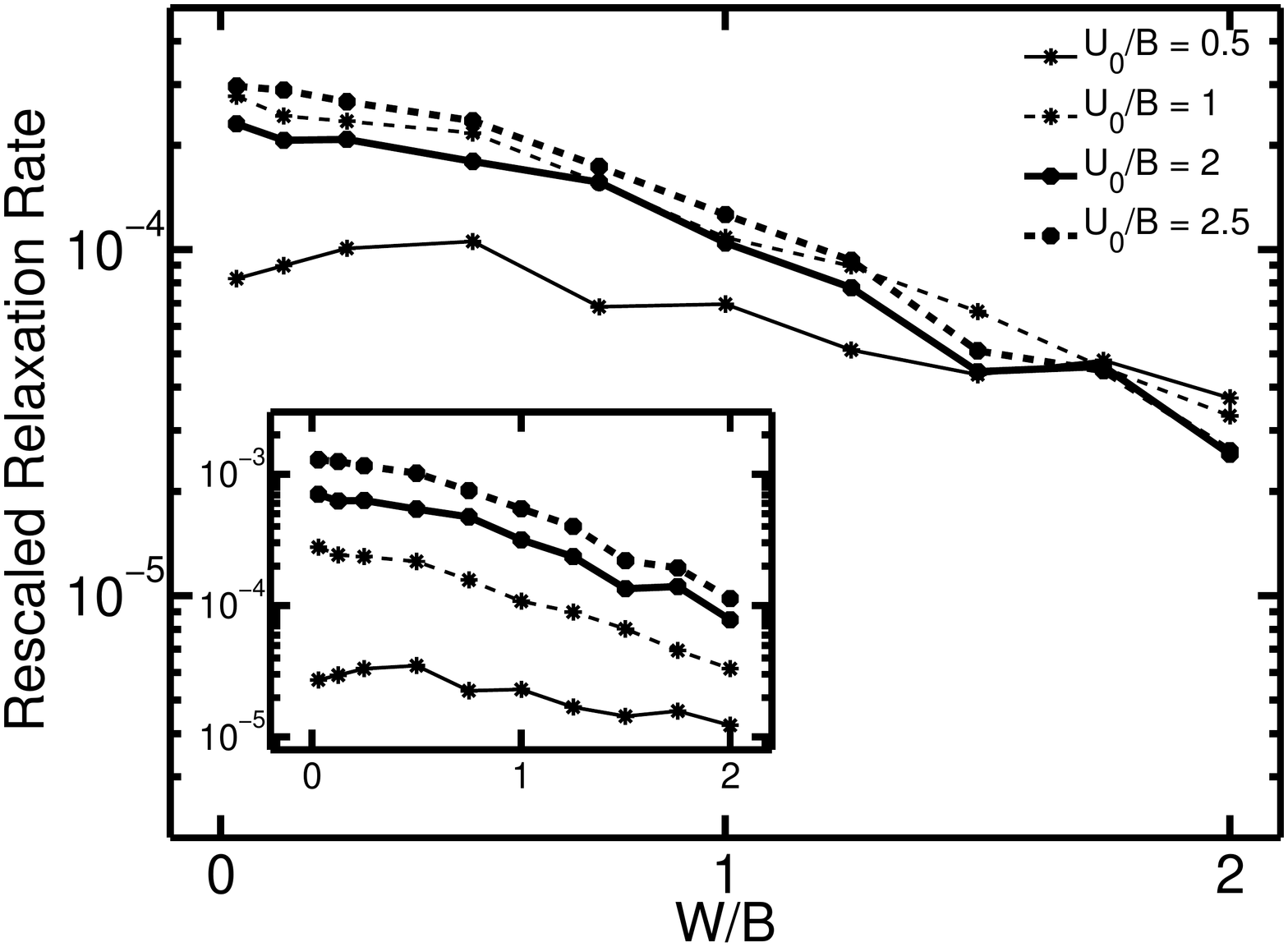}
  \caption{Relaxation Rate $\Gamma$ as a function of disorder strength for
  various interaction strengths. Data after rescaling as $(U_0/B)^c$, $c\approx 1.6$ is plotted in the main panel. The inset shows the raw data as a function of $W/B$.\label{fig:relaxRate}}
\end{figure}

Thus it is reasonable to determine $\Gamma$ for various disorder and
interaction strength values, both of which are measured in units of the
bare bandwidth, $B=4J$. The results of these calculations are shown in
\reff{fig:relaxRate}, where $\Gamma$ is plotted as a function of disorder
strength. The different lines belong to different interaction strengths.
The {rate} $\Gamma$ is also given in units {of} $B$.

The inset in \reff{fig:relaxRate} contains the raw data which after a
rescaling as $(U_0/B)^c$ is shown in the main panel, where $c\approx 1.6$.
A similar scaling of the relaxation rate was found in \cite{Varga03},
however with an exponent closer to the value of 2, but more work {needs to}
be done to understand the nature of these scaling exponents. Clearly in
\reff{fig:sigma} the case with low interaction strength deviates from the
major trend substantially.

\section{Summary}

In the present work we have shown that the interplay of both the disorder
and the long-range interaction can be investigated on the same footing in
order to study the relaxation process of an initially non-equilibrium
one-particle occupation distribution towards an equilibrium one. The
numerical solution of the equation of motion of the density matrix yields
an exponential form for the deviation of the distribution from a
Fermi-Dirac one. Additionally we found that the total single-particle
energy gradually decreases as a function of time. This way the relaxation
process in a completely closed system is possible via the rearrangement of
the energy into other components of the otherwise constant total energy. A
similar result has been found in \cite{Varga03}, which, however, contained
a phenomenological coupling to an environment.
{Morevoer,} in the present calculation the
off-diagonal matrix elements of the density matrix play an important role.

The results indicate that further advances can be achieved by
improving the present method. In a subsequent work the effect of higher
order correlations will be investigated, as well as the case of short
ranged interaction.

\begin{acknowledgement}

P.\ B.\ and H.\ S.\ gratefully acknowledge the financial support by the
European Commission, Marie Curie Excellence Grant MEXT-CT-2005-023778
(Nano\-electro\-photonics). I.V. thanks for financial support from
OTKA (Hungarian Research Fund) under Contracts No.\ T46303 and from
European Commission Contract No. MRTN-CT-2003-504574.

\end{acknowledgement}

% \newpage
\def\bstname{pss}


\begin{thebibliography}{[1]}

\bibitem{reviews}F. Evers and A.D. Mirlin, {\tt arXiv:0707.4378};
C. DiCastro and R. Raimondi, in: Proceedings of the International School
of Physics "Enrico Fermi", Varenna, Italy, 2003, (IOS Press, Bologna, 2004),
pp. 259-333.

\bibitem{Basko}D.~M.~Basko, I.~L.~Aleiner, and B.~L.~Altshuler, Ann. Phys.
\textbf{321}, 1126 (2006).

\bibitem{Gersh}I.~L.~Aleiner, B.~L.~Altshuler, and M.~E.~Gershenson, Waves in
Rand. Med. \textbf{9}, 201 (1999).

\bibitem{PQE} M.~Kira, F.~Jahnke, W.~Hoyer, and S.~W.~Koch,
Progress in Quantum Electronic, \textbf{23}, No.~6, (1999).

\bibitem{CSO}T.~Meier, P.~Thomas, and S.~W.~Koch,
Coherent Semiconductor Optics: From Basic Concepts to
Nanostructure Applications (Springer-Verlag, Berlin, 2007).

\bibitem{Bozsoki:PRL06}{P.~Bozsoki \textit{et~al.}, Phys.~Rev.~Letters, \textbf{97},
227402 (2006), {\tt arXiv:cond-mat/0611411}.}

\bibitem{Bozsoki:JLumi07}{P.~Bozsoki \textit{et~al.}, Journal of Luminescence, \textbf{124}, 99
(2007), {\tt arXiv:cond-mat/0505207}.}

\bibitem{Varga03}I.~Varga \textit{et~al.},
Phys.~Rev.~B, \textbf{68}, 113104 (2003), {\tt arXiv:cond-mat/0211206}.

\end{thebibliography}
\end{document}